# USING SECOND-ORDER HIDDEN MARKOV MODEL TO IMPROVE SPEAKER IDENTIFICATION RECOGNITION PERFORMANCE UNDER NEUTRAL CONDITION

*Ismail Shahin*

Electrical/Electronics and Computer Engineering Department
University of Sharjah, P. O. Box 27272, Sharjah, United Arab Emirates
E-mail: ismail@sharjah.ac.ae

## ABSTRACT

In this paper, second-order hidden Markov model (HMM2) has been used and implemented to improve the recognition performance of text-dependent speaker identification systems under neutral talking condition. Our results show that HMM2 improves the recognition performance under neutral talking condition compared to the first-order hidden Markov model (HMM1). The recognition performance has been improved by 9%.

## 1. INTRODUCTION

The use of hidden Markov model (HMM) for speech and speaker recognition has become popular in the last two decades. HMM has become one of the most successful and broadly used modeling technique for speech and speaker recognition [1].

HMM uses Markov chain to model the changing statistical characteristics that exist in the actual observations of speech signals. The Markov process is a double stochastic process where there is an unobservable Markov chain defined by a state transition matrix, and where each state of the Markov chain is associated with either a discrete output probability distribution (discrete HMM) or a continuous output probability density function (continuous HMM) [2]. The theory of Markov chains has been applied to speech and speaker recognition [3].

HMM is a powerful technique in optimizing the parameters that are used in modeling the speech signals. This optimization decreases the computational complexity in the decoding procedure and improves the recognition accuracy [2].

In the last two decades, the majority of the work performed in the field of speech and speaker recognition on hidden Markov model (HMM) has been done using HMM1 [2-7].

In this paper, HMM2 is being used and implemented to train and test text-dependent speaker identification systems under neutral talking condition.

## 2. SECOND-ORDER HIDDEN MARKOV MODEL

In HMM1, the underlying state sequence is a first-order Markov chain where the stochastic process is specified by a 2-D matrix of a priori transition probabilities ($a_{ij}$) between states $s_i$ and $s_j$, while in HMM2 the underlying state sequence is a second-order Markov chain where the stochastic process is specified by a 3-D matrix ($a_{ijk}$) [8,9]:

$$a_{ijk} = \text{Prob}\left(q_t = s_k \mid q_{t-1} = s_j, q_{t-2} = s_i\right) \quad (1)$$

with the constraints:

$$\sum_{k=1}^{N} a_{ijk} = 1 \qquad N \geq i, j \geq 1$$

where $N$ is the number of states in the model, and $q_t$ is the actual state at time $t$.

The probability of the state sequence, $Q \triangleq q_1, q_2, ..., q_T$ is defined as:

$$\text{Prob}(Q) = \Pi_{q_1} a_{q_1 q_2} \prod_{t=3}^{T} a_{q_{t-2} q_{t-1} q_t} \quad (2)$$

where $\Pi_i$ is the probability of state $s_i$ at time $t = 1$, $a_{ij}$ is the probability of the transition from state $s_i$ to state $s_j$ at time $t = 2$, and $T$ is the utterance length.

Each state $s_i$ is associated with a mixture of Gaussian distributions:

$$b_i(O_t) \triangleq \sum_{m=1}^{M} c_{im} N(O_t; \mu_{im}, \Sigma_{im}),$$

$$\text{with } \sum_{m=1}^{M} c_{im} = 1 \quad (3)$$

where the vector $O_t$ is the input vector at time $t$.

Given a sequence of observed vectors, $O \triangleq O_1, O_2, ..., O_T$, the joint state-output probability is defined as:

$$\text{Prob}(Q, O | \lambda) = \Pi_{q_1} b_{q_1}(O_1) a_{q_1 q_2} b_{q_2}(O_2) \cdot \prod_{t=3}^{T} a_{q_{t-2} q_{t-1} q_t} b_{q_t}(O_t) \quad (4)$$

## 3. EXTENDED VITERBI AND BAUM-WELCH ALGORITHMS

The most likely state sequence can be found by using the probability of the partial alignment ending at transition $(s_j, s_k)$ at times $(t-1, t)$:

$$\delta_t(j, k) \triangleq$$
$$\text{Prob}(q_1, ..., q_{t-1} = s_j, q_t = s_k, O_1, O_2, ..., O_t | \lambda)$$
$$T \geq t \geq 2, N \geq j, k \geq 1 \quad (5)$$

Recursive computation of the forward function $\alpha_t(j, k)$ is given by:

$$\delta_t(j, k) = \max_{N \geq i \geq 1} \{\delta_{t-1}(i, j) \cdot a_{ijk}\} \cdot b_k(O_t)$$
$$T \geq t \geq 3, N \geq j, k \geq 1 \quad (6)$$

The forward function $\alpha_t(j, k)$ defines the probability of the partial observation sequence, $O_1, O_2, ..., O_t$, and the transition $(s_j, s_k)$ between time $t-1$ and $t$:

$$\alpha_t(j, k) \triangleq \text{Prob}(O_1, ..., O_t, q_{t-1} = s_j, q_t = s_k | \lambda)$$
$$T \geq t \geq 2, N \geq j, k \geq 1 \quad (7)$$

$\alpha_t(j, k)$ can be computed from the two transitions $(s_i, s_j)$ and $(s_j, s_k)$ between states $s_i$ and $s_k$:

$$\alpha_{t+1}(j, k) = \sum_{i=1}^{N} \alpha_t(i, j) \cdot a_{ijk} \cdot b_k(O_{t+1})$$
$$T - 1 \geq t \geq 2, N \geq j, k \geq 1 \quad (8)$$

The backward function $\beta_t(i, j)$ can be expressed as:

$$\beta_t(i, j) \triangleq \text{Prob}(O_{t+1}, ..., O_T | q_{t-1} = s_i, q_t = s_j, \lambda)$$
$$T - 1 \geq t \geq 2, N \geq i, j \geq 1 \quad (9)$$

where, $\beta_t(i, j)$ is defined as the probability of the partial observation sequence from $t+1$ to $T$, given the model $\lambda$ and the transition $(s_i, s_j)$ between times $t-1$ and $t$.

## 4. SPEECH DATA BASE

In this research, our speech data base consists of twenty different speakers (ten adult males and ten adult females) uttering the same utterance (10 different isolated-word utterances) nine times under neutral talking condition.

Our speech data base is captured by a speech acquisition board using a 10-bit A/D converter and sampled at a sampling rate of 8 KHz. The speech data base is then preprocessed in order to be treated and analyzed easily.

## 5. RESULTS

In this research, linear predictive coding (LPC) cepstral feature analysis is used to form the observation vector for HMM1 and HMM2 (left-to-right model in both HMM1 and HMM2). Cepstral based features have been used extensively in speech recognition applications because they have been shown to outperform linear predictive coefficients. Cepstral based features attempt to incorporate the nonlinear filtering characteristics of the human auditory system in the measurement of spectral band energies.

The number of states, $N$, is equal to 5, the number of mixtures, $M$, is equal to 5 per state, with continuous mixture observation density is selected in HMM1 and HMM2.

Our results show that using HMM2 in the training and testing phases of text-dependent speaker identification systems under neutral talking condition improves the recognition performance compared to that using HMM1. The recognition performance has been improved by 9%.

Table 1 compares the results of the recognition performance of text-dependent speaker identification systems under neutral talking condition when HMM2 is used with that when HMM1 is used.

**Table 1.** Recognition performance under neutral talking condition using HMM1 and HMM2

| Model | Recognition performance |
|-------|------------------------|
| HMM1  | 90%                    |
| HMM2  | 98%                    |

## 6. DISCUSSION AND CONCLUSIONS

In this work, we showed that HMM2 yields high speaker identification recognition performance under neutral talking condition. Using HMM2 in the training and testing phases of text-dependent speaker identification systems under neutral talking condition improves the recognition performance compared to that using HMM1. The recognition performance has been improved by 9%.

Despite the success of using HMM1 under neutral talking condition, HMM2 is more appropriate model than HMM1 for text-dependent speaker identification systems because the underlying state sequence in HMM2 is a second-order Markov chain where the stochastic process is specified by a 3-D matrix, while in HMM1 the underlying state sequence is a first-order Markov chain where the stochastic process is specified by a 2-D matrix.

A naïve implementation of the recursion for the computation of $\alpha$ and $\beta$ in HMM2 requires on the order of $N^3T$ operations, compared with $N^2T$ operations in HMM1. Therefore, more memory space is required in HMM2 than that in HMM1.